\documentclass[12pt,a4paper]{revtex4}
\usepackage{graphicx}
\usepackage{amsmath}
\usepackage[usenames,dvipsnames]{color}
\usepackage{bm}
\usepackage[sort&compress]{natbib}

\begin{document}

\title{Synthesis of a fullerene-based one-dimensional nanopolymer through topochemical transformation of the parent nanowire }

\author{Junfeng Geng$^{1}$}
\email{jg201@cam.ac.uk, ilia@fias.uni-frankfurt.de}
\author{Ilia A Solov'yov$^{2,3,{\rm *}}$}
\author{David G. Reid$^{1}$, Paul Skelton$^{1}$, Andrew E. H. Wheatley$^{1}$, Andrey V. Solov'yov$^{2,3}$ and Brian F. G. Johnson$^{1}$}

\address{$^{1}$Department of Chemistry, University of Cambridge, Cambridge, CB2 1EW (United Kingdom)}
\address{$^{2}$Frankfurt Institute for Advanced Studies, Ruth-Moufang-Str. 1, 60438 Frankfurt (Germany)}
\address{$^{3}$A. F. Ioffe Physical-Technical Institute, Politechnicheskaya 26, 194021 St. Petersburg, (Russia)}

\begin{abstract}
Large-scale practical applications of fullerene (C$_{60}$) in nanodevices could be significantly facilitated if the commercially-available micrometer-scale raw C$_{60}$ powder were further processed into a one-dimensional (1D) nanowire-related polymer displaying covalent bonding as molecular interlinks and resembling traditional important conjugated polymers. However, there has been little study thus far in this area despite the abundant literature on fullerene. Here we report the synthesis and characterization of such a C$_{60}$-based nanowire polymer, (-C$_{60}$TMB-)$_n$, where TMB=1,2,4-trimethylbenzene, which displays a well-defined crystalline structure, exceptionally large length-to-width ratio and excellent thermal stability. The material is prepared by first growing the corresponding nanowire through a solution phase of C$_{60}$ followed by a topochemical polymerization reaction in the solid state. Gas chromatography, mass spectrometry and $^{13}$C nuclear magnetic resonance evidence is provided for the nature of the covalent bonding mode adopted by the polymeric chains. Theoretical analysis based on detailed calculations of the reaction energetics and structural analysis provides an in-depth understanding of the polymerization pathway. The nanopolymer promises important applications in biological fields and in the development of optical, electrical, and magnetic nanodevices.
\end{abstract}

\maketitle

\section{Introduction}

Two important forms of carbon, fullerene (C$_{60}$) and carbon nanotubes, are closely related to each other by their structural commonality of the sp$^2$ framework. Carbon nanotubes have been widely investigated for the last decade or so as one-dimensional (1D) nanomaterial \cite{ref1}. In contrast, fullerene 1D nanostructures, such as C$_{60}$ nanowires, presently represent laboratory curiosities \cite{ref2,ref3,ref4,ref5}. C$_{60}$ nanowires are highly interesting nanomaterials because of the physical properties associated with their high surface area, low-dimensionality and quantum confinement effects \cite{ref6,ref7}. In comparison with carbon nanotubes, however, C$_{60}$ nanowires are mechanically fragile and electrically non-conducting on account of the large intermolecular spaces in the crystalline lattice although an individual C$_{60}$ molecule is extremely hard, almost incompressible and electronically fully conjugated across all the carbon atoms \cite{ref6,ref8}.

It is expected that, in contrast to their parent nanowires or individual fullerene building blocks, polymerized C$_{60}$ nanowires will exhibit remarkably different and/or improved physical properties owing to the formation of polymeric chains and interchain cross-linking networks. From an applications point of view, a major advantage of using such polyfullerenes would be their bio-compatibility as they are totally free of metal. Notably, this clearly contrasts to the case of carbon nanotubes, the growth of which is catalyzed by metal nanoparticles, and from which by no means all the metal can be removed by post-purification processes \cite{ref9}. In this respect, C$_{60}$ nanowire polymers could be biologically even more attractive than carbon nanotubes for uses in, for example, drug delivery and biosensor preparation \cite{ref10,ref11}. Additionally, 1D C$_{60}$ polymers are conjugated or nearly-conjugated materials that promise outstanding photo-electrical properties because of the large magnitude of the nonlinear optical response of C$_{60}$ and its excellent photoinduced charge transfer behaviour \cite{ref6}. As such, they promise applications as optical limiters and photoconductors for solar energy devices, fuel cells, and field-emission transistors \cite{ref12,ref13,ref14,ref15,ref16}. All this makes them important materials for further study.

Fundamental to the development of applications is the question of how to make such metal-free, C$_{60}$-based nanowire polymers? In seeking to answer this question, we note first that a limited number of polymerization reactions of C$_{60}$ have been reported in literature \cite{ref17,ref18,ref19,ref20,ref21,ref22}. For example, it has been shown that the polymerization of thin solid films of C$_{60}$ may occur upon irradiation with visible-UV light.  In this case, the thin-films were prepared by vacuum deposition of pure C$_{60}$ on a substrate \cite{ref17}. On the basis of Raman scattering studies, such polymerization has been hypothesized to result from a [2+2] cycloadduction linkage between adjacent C$_{60}$ molecules in the structure \cite{ref17,ref18}. Single crystals of KC$_{60}$, prepared by the co-evaporation of K and C$_{60}$, have undergone polymerization via formation of the same [2+2] structure motif, but the whole polymer is a linear polyanion network, as suggested by electron spin resonance (ESR) measurements, pointing to a possible donation of electrons from K atoms to the polymer chains \cite{ref19}. An in-chain metal-fullerene (-PdC$_{60}$-)$_n$ polymer, synthesized by the reaction of C$_{60}$ with a palladium complex, Pd$_2$[(C$_6$H$_5$CH=CH)$_2$C=O] CHCl$_3$, has been noted \cite{ref20}. Moreover, side-chain C$_{60}$ copolymers formed through the attachment of C$_{60}$ to amino polymer substrate have also been reported, with the coupling of C$_{60}$ to the amino polymer being analogous to that observed for amines \cite{ref21,ref22}.

It has also been demonstrated that crystallization of C$_{60}$ molecules from organic solvents yields C$_{60}$ nanocrystals with varied shapes and structures, and that these shapes and structures are dependent upon the crystallization conditions and the solvent(s) employed \cite{ref23,ref24,ref25}. One example is the preparation of C$_{60}$ nanowhiskers with a claimed face-centered cubic structure using a liquid-liquid interface method \cite{ref26}. This method has also been extended to prepare the nanowhiskers of C$_{70}$, iodine-doped C$_{60}$, and the C$_{60}$-based complex of ($\eta^2$-C$_{60}$)Pt(PPh$_3$)$_2$ \cite{ref26,ref28}. In the second example, crystallization of C$_{60}$ molecules from a 1,2,4-trimethylbenzene (1,2,4-TMB) solution of C$_{60}$ has been found to be able to produce exceptionally long C$_{60}$ nanowires under appropriate conditions, with a length-to-width aspect ratio as large as 3000, and an orthorhombic nanostructure \cite{ref29,ref30}.

In spite of these literature reports, it is clear that there has been little study in the related field of nanowire-oriented C$_{60}$ polymer. This new class of 1D nanomaterial exhibits similarities in structure to other important conjugated polymers such as polyacetylenes, polyanilines, and polyphenylenes. Here we report the synthesis and characterization of such a C$_{60}$-based nanowire polymer, (-C$_{60}$TMB-)$_{n}$, which is formed by the polymerization occurring in their parent nanowire via a topochemical solid-state reaction. Because the reactive monomers are pre-organized at a distance commensurate with the repeat distance in the final polymer, the application of thermal or photochemical energies to the solid induces polymerization. In order to understand the polymerization pathway, we have employed gas chromatography (GC), mass spectrometry (MS) and $^{13}$C nuclear magnetic resonance (NMR) spectroscopy to investigate the nature of the bonds formed during the polymerization process. Additionally, to explain the polymerization mechanism in more detail we have theoretically analyzed the associated molecular orientations, structural arrangements and the favorable reaction pathways.

\section{Experimental Details}

GC-MS analysis was carried out using a Perkin Elmer Turbomass GC-MS. This system utilizes an Autosystem XL GC connected to a small quadrupole mass analyzer. Samples were dissolved in CCl$_4$ prior to analysis, and were introduced to the system using a split injection technique. After a delay of 1 minute the mass spectrometer scan was initiated with the parameters set out below.

GC conditions: Injection volume, 0.5 $\mu$l; Injection temperature, 200 $^{\circ}$C; Helium carrier gas pressure, 40 psi. Column: Perkin Elmer Elite PE-5MS (5 \% phenyl, 95 \% methyl polysiloxane); Length, 30 m; ID, 0.25 mm; film thickness, 0.25 $\mu$m. Column gradient: Start temperature, 60 $^{\circ}$C, for 1 minute then ramped at 10 $^{\circ}$C/minute to 200 $^{\circ}$C, which temperature was maintained for 2 minutes. Total analysis time was 17 minutes.

MS Spectrometer conditions: Ionization mode, EI+ ; Ionization energy, 70 eV ; Source temperature, 200 $^{\circ}$C ; Transfer line temperature, 200 $^{\circ}$C; Mass range scanned, 50-650 Da; Scan time, 0.5 s; Interscan time, 0.1 s; Solvent delay time, 1 minute.

$^{13}$C NMR spectra were acquired at 100.62 MHz on Bruker AVANCE 400 standard bore (liquid state) and wide bore (solid state) spectrometers with standard Bruker broadband tuneable probes. Samples for solid state NMR were packed into 4 mm zirconia rotors padded with polytetrafluoroethylene tape to take up excess space due to limited sample volumes. Relevant spectral acquisition parameters were as below.

Solution-state NMR spectra were obtained using pulse-acquire with 10  s  $\pi$/2 pulses, a recycle delay of 5 s, and broadband $^1$H decoupling, and referenced relative to tetramethylsilane at 0 ppm.  Solid-state NMR spectra were acquired with standard cross polarization magic angle spinning (CP-MAS) (12.5 kHz, 2.5 $\mu$s $^1$H pulse, ramped CP time 5 ms, recycle delay 1 s) and high power broadband 1H decoupling during acquisition (fullerene nanowires).  Pure fullerene was observed using direct polarization MAS (5 ms $^{13}$C $\pi$/2 pulse), recycle delay 60 s). Chemical shifts are referenced to the methylene carbon signal of glycine at 43.1 ppm which is back referenced to tetramethylsilane.

Transmission electron microscopy (TEM) and selected area electron diffraction (SAED) were performed using a JEOL JEM-3010x microscope operated at 300 kV and a JEOL 200CXi microscope operated at 200 kV, respectively. Specimens were directly deposited on copper grids coated with holey carbon films without any prior suspension treatment in solution.

Scanning electron microscopic (SEM) examinations were performed using a LEO-32 electron microscope operated at 5 kV. Samples were directly deposited on a specimen holder (carbon mat) without surface coating of a conducting material.

Raman spectroscopic studies were performed using a micro-Raman spectrometer with incident Ar ion laser at   $\lambda$=514.5 nm and the Renishaw software. Extra care was taken to avoid laser damage to the samples. To ensure this, the laser power employed was kept as low as possible, and optical microscopic examinations were carried out for the nanowires before and after each measurement.

\section{Theoretical Methods}
\label{sec3}

The structures and properties of the molecules involved in the nanowire polymerization process were studied theoretically. The calculations were performed with the Gaussian 03 software package \cite{ref31}, within the framework of the {\it ab initio} density functional theory (DFT). Within the framework of the DFT one has to solve the Kohn-Sham equations \cite{ref32,ref33,ref34}, which read as (in the atomic system of units):

\begin{equation}
\left( {\frac{{\hat p^2 }}{2} + \hat U_{ions}  + \hat V_H  + \hat V_{xc} } \right)\psi _i  = \varepsilon _i \psi _i
\label{eq1}
\end{equation}

\noindent
Here the first term represents the kinetic energy of the i-th electron, and $\hat U_{ions}$ describes its attraction to the ions in the system, while $\hat V_H$  is the Hartree part of the interelectronic interaction:

\begin{equation}
\hat V_H (\vec r) = \int {\frac{{\rho (\vec r')}}{{\left| {\vec r - \vec r'} \right|}}} d\vec r'
\label{eq2}
\end{equation}

\noindent
and $\rho (\vec r')$  is the electron density:

\begin{equation}
\rho (\vec r) = \sum\limits_{\nu  = 1}^N {\left| {\psi _i (\vec r)} \right|} ^2
\label{eq3}
\end{equation}

\noindent
$\hat V_{xc}$ in Eq.~(\ref{eq1}) is the local exchange-correlation potential, $\psi_i$ are the electronic orbitals and $N$ is the number of electrons in the system. The exchange-correlation potential is defined as the functional derivative of the exchange-correlation energy functional:

\begin{equation}
V_{xc}  = \frac{{\delta E_{xc} \left[ \rho  \right]}}{{\delta \rho (\vec r)}}
\label{eq4}
\end{equation}

\noindent
The functionals employed by DFT methods partition the exchange-correlation energy into exchange and correlation parts. The local exchange functional is virtually always defined as follows:

\begin{equation}
E_x^{LDA}  =  - \frac{3}{2}\left( {\frac{3}{{4\pi }}} \right)^{1/3} \int \rho  ^{4/3} d^3 \vec r
\label{eq5}
\end{equation}

\noindent
This form was developed to reproduce the exchange energy of a uniform electron gas. By itself, however, it is not sufficient for an adequate description of a many-body system. A more accurate parametrization of the exchange functional introduced by Becke \cite{ref35}, which includes gradient-corrections and reads as:

\begin{equation}
E_x^{B88}  = E_x^{LDA}  - \gamma \int {\frac{{\rho ^{4/3} x^2 }}{{1 + 6\gamma \sinh ^{ - 1} x}}} d^3 \vec r
\label{eq6}
\end{equation}

\noindent
where $x = \rho ^{ - 4/3} \left| {\nabla \rho } \right|$ and $\gamma  = 0.0042$ is a parameter chosen to fit the known exchange energies of the noble gas atoms.

In spite of the success of pure DFT theory in many cases, one has to admit that the Hartree-Fock theory accounts for the electron exchange more naturally and precisely. Thus, Becke has suggested hybrid functionals \cite{ref35} which include a mixture of Hartree-Fock and DFT exchange along with DFT correlations:

\begin{equation}
E_{xc}^{mix}  = c_{HF} E_x^{HF}  + c_{DFT} E_{xc}^{DFT}
\label{eq7}
\end{equation}

\noindent
where $c_{HF}$ and $c_{DFT}$ are constants. Following this idea, a Becke-type three parameter functional (B3LYP) can be defined as follows:

\begin{eqnarray}
\nonumber
E_{xc}^{B3LYP}  &=& E_x^{LDA}  + c_0 \left( {E_x^{HF}  - E_x^{LDA} } \right) + c_x \left( {E_x^{B88}  - E_x^{LDA} } \right) +\\
 &&E_c^{VWN3}  + c_c \left( {E_c^{LYP}  - E_c^{VWN3} } \right)
\label{eq8}
\end{eqnarray}

\noindent
Here, $c_0=0.2$, $c_x=0.72$ and $c_c=0.81$ are constants, which were derived by fitting to the atomization energies, ionization potentials, proton affinities and first-row atomic energies \cite{ref31,ref36}. $E_x^{LDA}$ and $E_x^{B88}$ are defined in Eqs.~(\ref{eq5}) and (\ref{eq6}) respectively. $E_x^{HF}$ is the functional corresponding to Hartree-Fock equations. The explicit form for the correlation functional $E_c^{VWN3}$ as well as for gradient-corrected correlation functional of Lee, Yang and Parr, $E_c^{LYP}$, can be found in Ref.~\cite{ref37} and Ref.~\cite{ref38} respectively. We have used the B3LYP functional for all computations in the present paper.

In Gaussian 03, the molecular orbitals, $\chi_i$, are approximated by a linear combination of a pre-defined set of single-electron functions, $\chi_{\mu}$, known as basis functions. This expansion reads as follows:

\begin{equation}
\psi _i  = \sum\limits_{\mu  = 1}^N {c_{\mu i} \chi _\mu  }
\label{eq9}
\end{equation}

\noindent
where coefficients $c_{\mu i}$ are the molecular orbital expansion coefficients, and $N$ is the number of basis functions, which are chosen to be normalized.

The basis functions $\chi _\mu$ are defined as linear combinations of primitive gaussians:

\begin{equation}
\chi _\mu   = \sum\limits_p {d_{\mu p} g_p }
\label{eq10}
\end{equation}

\noindent
where $d_{\mu p}$ are fixed constants within a given basis set, the primitive gaussians, $g_p= g(\alpha,\vec r)$, are the gaussian-type atomic functions (for details see Ref.~\cite{ref36}).

In the calculations we accounted for all electrons in the system, and employed the standard STO-3G, 6-21G and 6-31G(d) basis sets \cite{ref36}. The STO-3G and 6-21G basis sets were used for geometry optimization and energy calculation, while the 6-31G(d) was used for simulating the nuclear magnetic resonance properties.

We have calculated chemical shielding tensors for different molecules of interest and compared the results with experimental measurements. According to NMR theory, the chemical shielding tensor can be computed as the second derivative of the electronic energy $E$ with respect to the external magnetic field $\vec B$ and the nuclear magnetic moment of interest $\vec m^N$:

\begin{equation}
\sigma _{ji}^N  = \left( {\frac{{d^2 E}}{{dB_i dm_j^N }}} \right)_{\vec B,\vec m^N  = 0}
\label{eq11}
\end{equation}

\noindent
Indices $i$ and $j$ correspond to the $i$-th and $j$-th components of the vectors $\vec B$ and $\vec m^N$ respectively. The gauge-origin problem inherent to finite basis set calculations of NMR shieldings can be solved by using gauge-including atomic orbitals (GIAOs) \cite{ref39,ref40,ref41,ref42}. In the GIAO method the corresponding calculations are carried out with the explicitly field-dependent basis function defined as follows:

\begin{equation}
\chi _\mu   = \exp \left( { - \frac{i}{{2c}}\left( {\vec B \times \vec R_\mu  } \right) \cdot \vec r} \right)\chi _\mu  (0)
\label{eq12}
\end{equation}

\noindent
Here $\chi _\mu  (0)$  is the field independent basis function defined in Eq.~(\ref{eq10}), $\vec R_\mu$ denotes the center of the corresponding  $\chi _\mu$, $\vec r$ describes the electron coordinate, and $c$ is the speed of light.

The chemical shielding tensor $\sigma _{ji}$ in Eq.~(\ref{eq11}) can be diagonalized to yield a tensor with three principal components, $\sigma _{11}  \le \sigma _{22}  \le \sigma _{33}$, which define the isotropic shielding as follows:

\begin{equation}
\sigma _{iso}  = \frac{{\sigma _{11}  + \sigma _{22}  + \sigma _{33} }}{3}
\label{eq13}
\end{equation}

\noindent
Isotropic shielding is an important quantity, because in a solution NMR experiment, rapid tumbling of the molecules commonly averages the chemical shielding tensor, and only the isotropic chemical shielding is detected. In solid samples, the presence of chemical shielding anisotropy often generates broad powder patterns, but in this case the magic-angle spinning (MAS) technique can be used to partially or completely average the chemical shielding anisotropy powder patterns.

Another important quantity is the chemical shift $\delta$, which is defined as the difference in shielding between the nucleus in the species under investigation, $\sigma _s$ , and the shielding of the same nucleus in a reference compound, $\sigma _{ref}$ :

\begin{equation}
\delta {\rm{ (ppm)}} = 10^6  \times \frac{{\sigma _{ref}  - \sigma _s }}{{1 - \sigma _{ref} }} \approx 10^6  \times \left( {\sigma _{ref}  - \sigma _s } \right)
\label{eq14}
\end{equation}

\noindent
The latter approximation is often used because usually  $\sigma _{ref}\ll1$. Equation (\ref{eq14}) gives the chemical shift in parts per million (ppm). In this work, we are interested in the chemical shifts for the $^{13}$C nuclei, which we have calculated relative to the chemical shift of the carbon atoms in tetramethylsilane (TMS).

\section{Results and Discussion}
\subsection{Synthesis and Characterization by GC-MS}

\begin{figure}[b]
\includegraphics[width=16cm,clip]{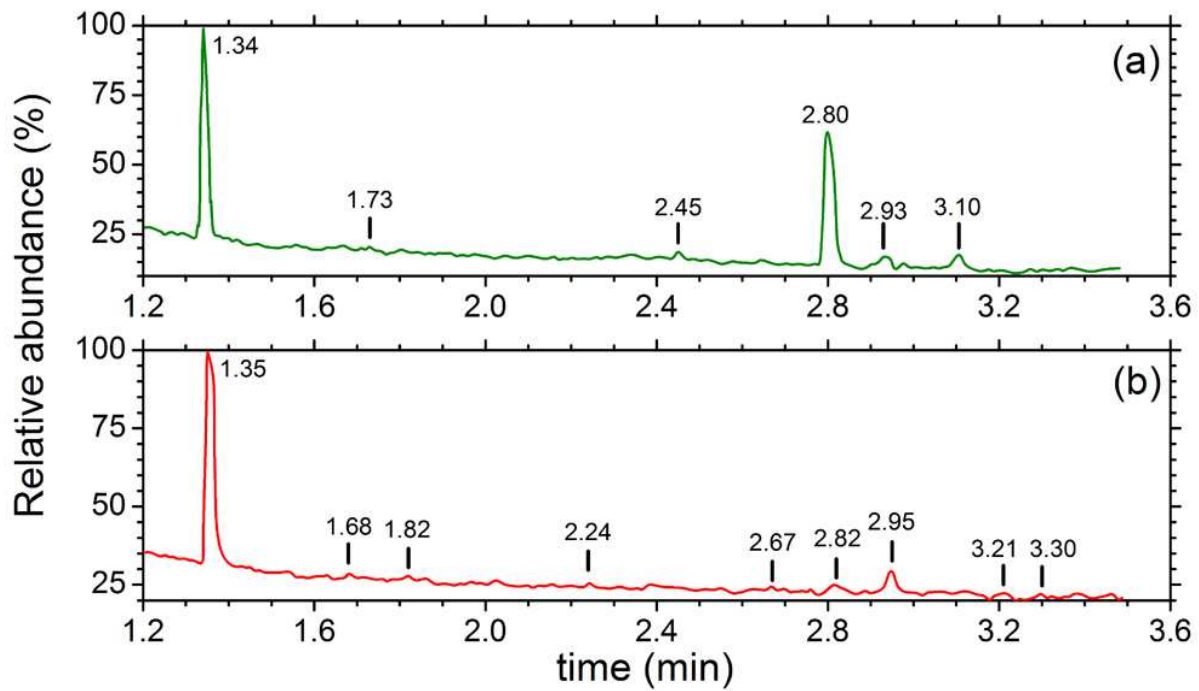}
\caption{GC analysis of the C$_{60}$TMB nanowires. (a) GC analysis of the as-prepared sample. (b) GC profile of the aged sample. The GC peaks at 1.34 minutes in (a) and 1.36 minutes in (b) are attributable to the remnant of CCl4 solvent, and the peaks at 2.93 or 2.95 minutes in the spectra are due to an impurity present in the system.}
\label{fig1}
\end{figure}

Synthesis of the fullerene nanowire polymer required that we first prepared C$_{60}$ nanowires using a 1,2,4-TMB solution of C$_{60}$ by a previously described method \cite{ref29}. We noted that the resulting nanowires were highly stable, as indicated by the fact that there was no detectable alteration in either their crystalline morphology, crystal color, or sample weight as a function of time. However, the solubility of the material was found to change with time. Unlike raw C$_{60}$ powder, well known to be highly soluble in aromatic solvents, the as-made nanowires were only partially soluble in these solvents, and this solubility decreased further with time. After ageing for a period of $\sim$10 months, the nanowires became totally insoluble in common organic solvents including benzene, toluene, 1,2,4-TMB, and carbon tetrachloride. This initial observation led us to speculate that a polymerization reaction might have occurred within the system, and that this had led to interesting new physical and chemical properties, including excellent thermal stability and solvent-resistant behavior.

\begin{figure}[t]
\includegraphics[width=16cm,clip]{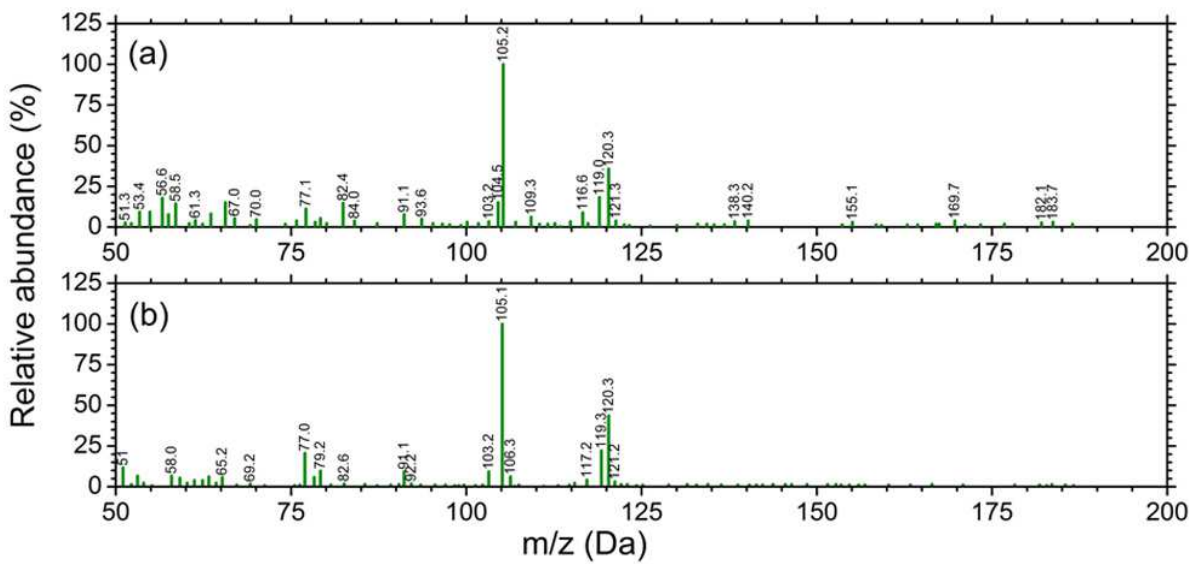}
\caption{MS analysis of C$_{60}$TMB nanowires. The profiles (a) and (b) correspond to the GC peaks at 2.80 and 3.10 minutes (elution times) as shown in Figure \ref{fig1}a.}
\label{fig2}
\end{figure}

Observation of the loss of solubility was consistent with the results obtained from the GC-MS analysis. Tests were carried out using two suspensions of the solid in CCl$_{4}$, corresponding respectively to as-made and aged nanowires. Prior to the tests, the solids were resident in the solvent for 5 days. The results are shown in Figure \ref{fig1}. As can be seen, GC successfully isolated 1,2,4-TMB molecules from the suspension containing the as-made nanowires, as indicated by the signals at 2.80 and 3.10 minutes (see Figure \ref{fig1}a). The mass spectra of these eluent fractions, shown in Figure \ref{fig2}, confirmed the presence of trimethylbenzene in both isomeric forms: the main GC peak (at 2.80 minutes) corresponded to 1,2,4-TMB, while the smaller peak (at 3.10 minutes) corresponded to either 1,3,5-TMB or 1,2,3-TMB, which existed as impurities in the solvent. The major ions in the mass spectra were seen at 120, 105, 91 and 77 Da and were consistent with those expected from trimethylbenzene. In contrast, no TMB molecules could be extracted from the aged solid, as confirmed by the absence of the corresponding GC peaks from the spectrum (see Figure \ref{fig1}b). These data support the observation of partial solubility behavior of the as-made solid, suggesting a possible partial polymerization in the sample, which appears to arise from the slow-growth of the nanowires. The data also suggest a likely full polymerization in the aged solid, wherein all the TMB molecules appear to have chemically bonded to C$_{60}$, resulting in an extensive, non-soluble polymeric network.

\subsection{$^{13}$C NMR Measurement and the Polymerization Bonding Mode}

\begin{figure}[b]
\includegraphics[width=16.4cm,clip]{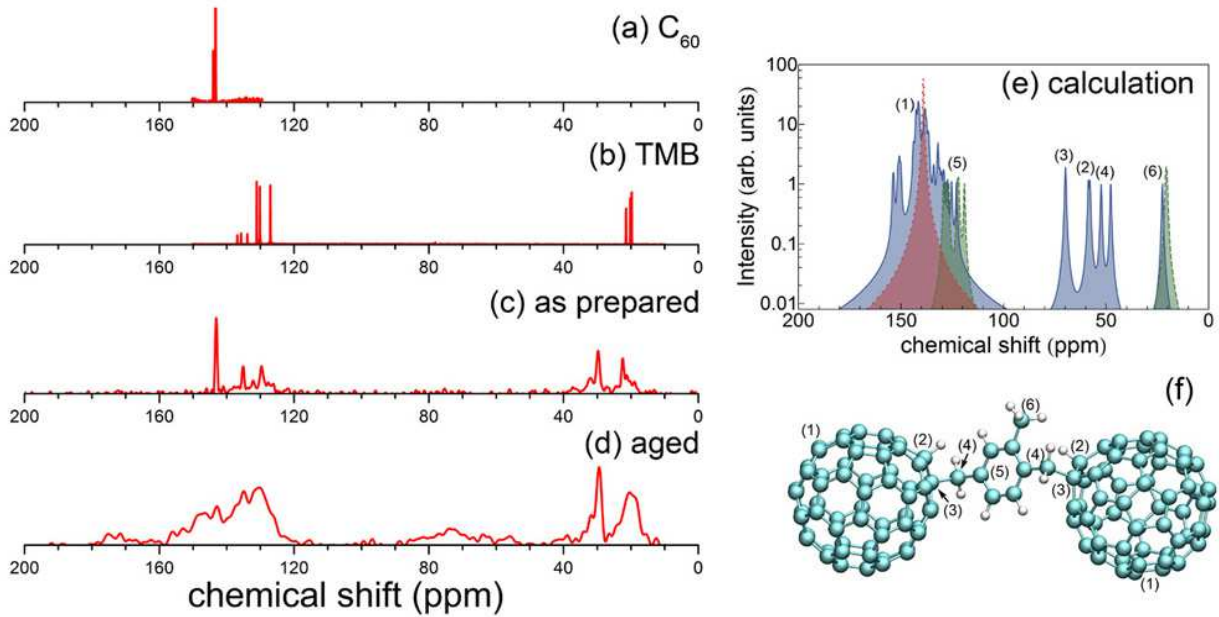}
\caption{(a) Direct polarization magic angle spinning (DP-MAS) $^{13}$C solid state spectrum of pure C$_{60}$. (b) Solution-state (in deuterochloroform) $^{13}$C NMR spectrum of pure 1,2,4-trimethylbenzene. (c) Solid-state CP-MAS $^{13}$C NMR spectrum of the as-prepared C$_{60}$-1,2,4-trimethylbenzene adduct. (d) Solid-state CP-MAS $^{13}$C NMR spectrum of the aged C$_{60}$-1,2,4-trimethylbenzene adduct. (e) The calculated spectra using the {\it ab initio} B3LYP/6-31G(d) method. Solid line (blue fill), dashed line (green fill) and dotted line (red fill) show the corresponding spectra for the C$_{60}$-TMB-C$_{60}$ complex as shown in (f), pure 1,2,4-trimethylbenzene and pristine C$_{60}$ respectively. The numbers labelled in (e) show the chemical shifts of the corresponding carbon atoms marked in (f).}
\label{fig3}
\end{figure}

In order to confirm the polymerization and elucidate the nature of the bonding mode involved, we performed $^{13}$C NMR spectroscopic studies on both as-made and aged specimens, and on raw C$_{60}$ powder (99.9 \%, SERS Ltd.) as well as 1,2,4-TMB solvent (98 \%, Aldrich) (see Figure \ref{fig3}). Some assignments are immediately suggested by a comparison with the spectra of the pure precursor materials, namely, the pristine C$_{60}$ (see Figure \ref{fig3}a) and pure 1,2,4-TMB (see Figure \ref{fig3}b). The broad peak at 22.5 ppm in the as-prepared C$_{60}$TMB adduct (see Figure \ref{fig3}c) results from the three methyl groups of 1,2,4-TMB (19.8, 20.2 and 21.5 ppm), the signal at 129.9 ppm in the same spectrum to protonated benzyl carbons of the adducted 1,2,4-TMB (c.f. signals at 127.0, 130.2 and 131.1 ppm in liquid 1,2,4-TMB), that at 135.2 ppm to the quaternary benzyl carbons (c.f. signals at 133.9, 135.7 and 136.8 ppm in liquid 1,2,4-TMB), and that at 143.3 ppm to the C$_{60}$ component of the adduct (c.f. signal at the same chemical shift in pure C$_{60}$). Apart from these, it is crucial to note the new peak exhibiting a chemical shift of 30.4 ppm in the spectrum. This is consistent with the presence of methylene carbon atoms directly bonded to aromatic rings \cite{ref43}, providing direct evidence for polymerization.

A comparison of the $^{13}$C NMR spectra of the as-prepared and the aged (see Figure \ref{fig3}d) materials clearly indicates that although signals in each sample are consistent with one another, in the latter case the peaks are much broader, consistent with the expected structural diversity and heterogeneity expected in the aged sample. In addition, a characteristic fingerprint of the spectrum of the aged nanowires is the new emergence of three broad peaks at $\sim$56 ppm, $\sim$64 ppm, and $\sim$73 ppm, respectively. These three peaks are consistent with what would be expected from the three carbons marked as 2, 3 and 4 in Figures \ref{fig3}e-f as the result of polymerization (also see the theoretical analysis part below), providing further evidence of polymerization.

\begin{table}[htbp]
\caption{$^{13}$C NMR spectroscopic data (ppm) of the molecules of C$_{60}$, 1,2,4-trimethylbenzene, 2,4,6,2$^{'}$,4$^{'}$,6$^{'}$-diphenylmethane and the (-C$_{60}$TMB-)$_{n}$ nanopolymer. The values in brackets indicate chemical shifts calculated in this work using the B3LYP/6-31G(d) method. The values in bold numbers are the chemical shifts measured in this work.
\emph{Notes: a: Reference \cite{ref45}; b: Reference \cite{ref44}; c: Reference \cite{ref43}; *: Quaternary carbons on the benzene ring.}
}
\label{tab1}
\begin{center}
\begin{ruledtabular}
\begin{tabular}{|r|cccc|}
  Carbon position & Pristine & 1,2,4-trimethylbenzene  & 2,4,6,2$^{'}$,4$^{'}$,6$^{'}$ & The (-C$_{60}$TMB-)$_{n}$ \\
                  & C$_{60}$ & (1,2,4-TMB)             & hexamethyl                    & nanopolymer \\
                  &          &                         & diphenylmethane               &              \\
  \hline
  C (sp$^2$ in C$_{60}$) & 142.7$^a$      &   &   & \textbf{143.3} \\
                         & \textbf{143.3} &   &   &  (130.3-154.0) \\
                         & (139.2)        &   &   &   \\
  Ring carbons: C-1      &                & 133.4$^{*b}$, \textbf{133.9}$^*$, (126.8) & 134.6$^c$     & (129.0) \\
                C-2      &                & 136.3$^{*b}$, \textbf{136.8}$^*$, (129.4) & 136.3$^{c*}$  & \textbf{135.2}$^{*}$, (132.0) \\
                C-3      &                & 130.5$^{b}$, \textbf{131.1}, (122.5)      & 129.0$^{c}$   & \textbf{129.9}, (127.4) \\
                C-4      &                & 135.2$^{*b}$, \textbf{135.7}$^*$, (128.1) & 134.5$^{c*}$  & (129.5) \\
                C-5      &                & 126.7$^{b}$, \textbf{127.0}, (119.1)      & 129.0$^{c}$   & (123.0) \\
                C-6      &                & 129.8$^{b}$, \textbf{130.2}, (121.9)      & 136.3$^{c*}$  & (125.4) \\
               1-CH$_3$  &                & 19.1$^{b}$, \textbf{19.8}, (20.3)         &               &  \\
               2-CH$_3$  &                & 19.5$^{b}$, \textbf{20.2}, (20.7)         & 20.9$^{c}$    &   \\
               3-CH$_3$  &                &                                           &               & \textbf{22.5}, (22.6) \\
               4-CH$_3$  &                & 20.9$^{b}$, \textbf{21.5}, (21.0)         & 20.9$^{c}$    &   \\
               5-CH$_3$  &                &                                           &               &  \\
               6-CH$_3$  &                &                                           & 20.9$^{c}$    &  \\
                -CH$_2$- &                &                                           & 31.2$^{c}$    & \textbf{30.4}, \textbf{55.9}  \\
                         &                &                                           &               & (28-43, 47.8, 52.4) \\
C (sp$^3$ in C$_{60}$)   &                &                                           &               & \textbf{73.2}, \textbf{63.5}  \\
                         &                &                                           &               & (69.6, 58.6) \\
\end{tabular}
\end{ruledtabular}
\end{center}
\end{table}

The $^{13}$C-NMR data indicate the plausibility of polymerization proceeding via conversion of the methyl carbons of TMB into methylenes when reacting with C$_{60}$ units to yield covalent C-C$_{60}$ bonds. Such a reaction logically leads to the formation of a polymethylated aromatic chain structure, i.e., a (-C$_{60}$TMB-)$_{n}$ type of copolymer. As such, the chemical shift of the methylene carbons in the polymeric structure, as observed at  $\delta$=30.4 ppm, is totally comparable to that of a structurally closely-related multi-substituted diphenylmethane such as 2,4,6,2$^{'}$,4$^{'}$,6$^{'}$-hexamethyl diphenylmethane ($\delta$=31.2 ppm) \cite{ref43,ref44,ref45}. A detailed comparison of the chemical shifts of the structurally related molecules, based on both experimental measurements and theoretical calculations, is presented in Table~\ref{tab1}.

\begin{figure}[htb]
\includegraphics[width=16cm,clip]{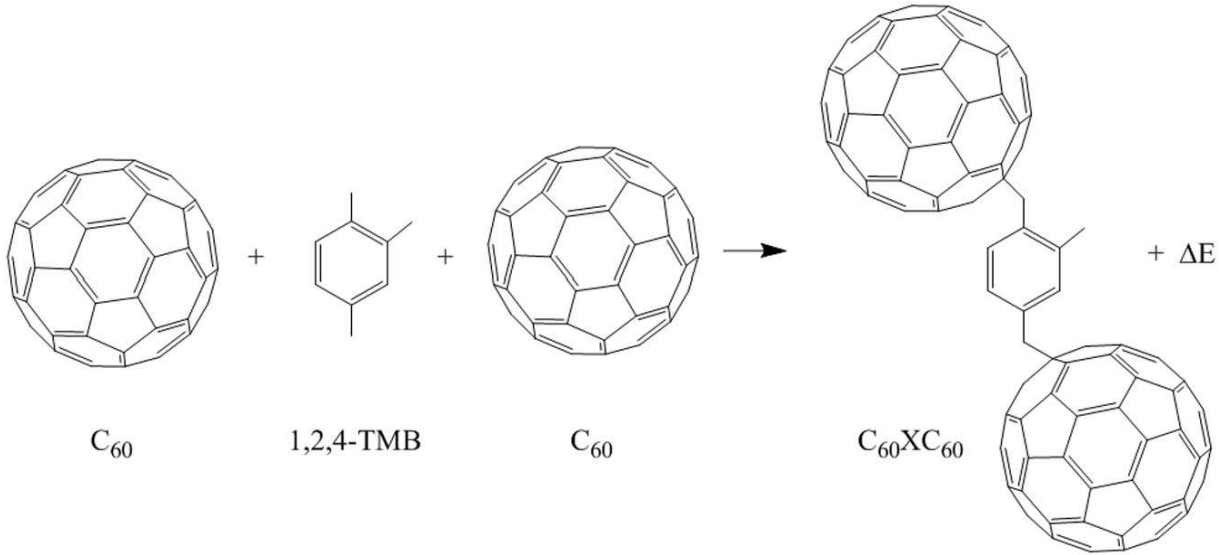}
\caption{A schematic shows the polymerisation reaction pathway. The label X in the product indicates a residue derived from the TMB molecule during the reaction. $\Delta E$ is the enthalpy of the reaction. Hydrogen atoms are omitted in this diagram for clarity.}
\label{fig4}
\end{figure}

A comparison of the $^{13}$C-NMR data with that obtained by GC-MS reveals that polymerization occurred following the nanowire growth, but that the reaction was not fully completed when the nanowire growth ceased. Thus, significant quantities of isolated TMB molecules remained unreacted in the crystalline lattice. These molecules could be easily extracted by dissolution of the crystal in a suitable organic solvent, as indicated by the GC-MS data. Evidently, polymerization continued in the nanowires until all TMB molecules were covalently bonded to adjacent fullerenes, at which point the extraction of further TMB molecules became impossible. On the basis of these analyses, a schematic illustrating the polymeric structure is shown in Figure \ref{fig4}.

\subsection{Spectroscopic Characterization by Laser-Raman, SEM, HRTEM and SAED}

\begin{figure}[!htb]
\includegraphics[width=14cm,clip]{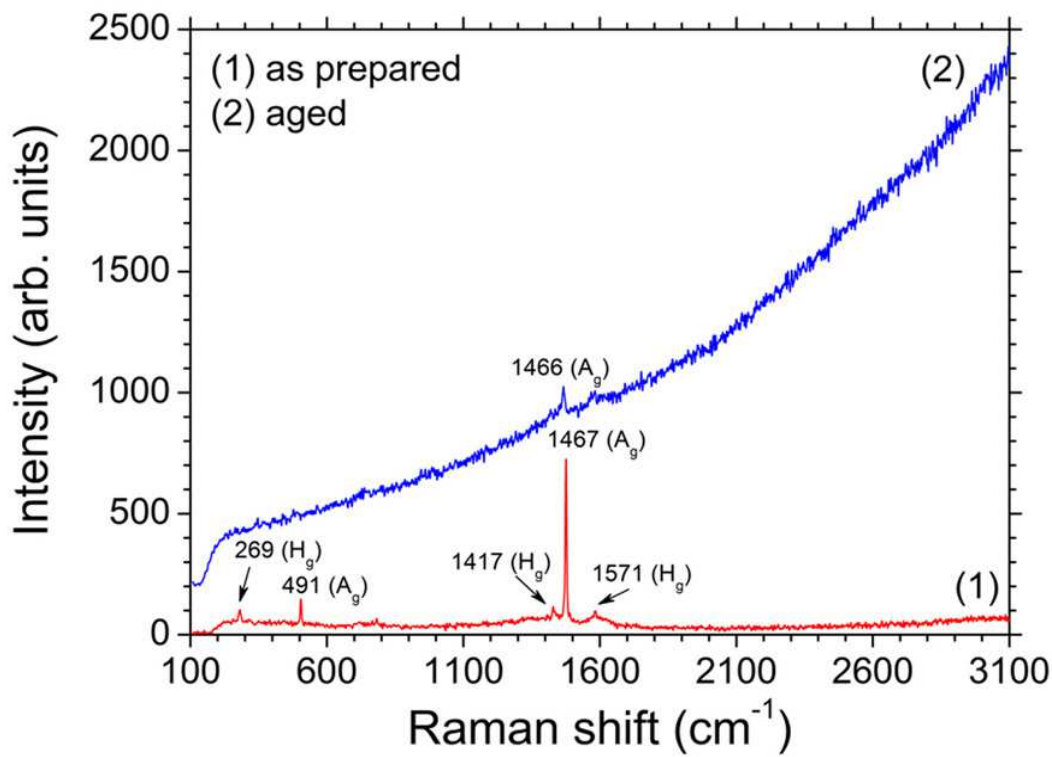}
\caption{Laser micro-Raman spectroscopic analysis shows the profiles of the as-made (1) and the aged (2) C$_{60}$TMB nanowire sample.}
\label{fig5}
\end{figure}
\begin{figure}[!htb]
\includegraphics[width=14cm,clip]{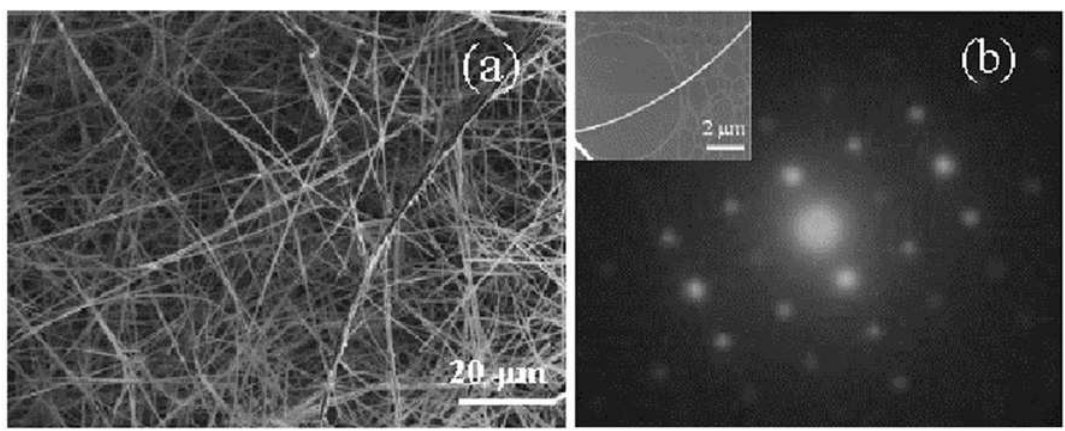}
\caption{(a) SEM image of the (-C$_{60}$TMB-)$_n$ nanopolymer. (b) SAED pattern shows the crystalline nature of the polymer. The inset in (b) is a typical TEM image of a nanopolymer associated with the SAED measurement.}
\label{fig6}
\end{figure}

Our observations suggest that the polymerization results from a process whereby nanowires undergo a chemical transformation in the crystalline lattice. The fixed lattice positions of the fullerene molecules render it straightforward for the guest TMB species to form bonds with them. Micro-Raman spectroscopic characterization (Ar$^{+}$ ion excitation laser,  $\lambda$=514 nm, spot size $\sim$ 10 $\mu$m) shows that, with the exception of the high-frequency and strongest Ag-symmetry `pentagonal-pinch' (pp) mode at $\sim$1468 cm$^{-1}$, the other Raman-active modes cannot be clearly identified in the polymerized sample (see Figure \ref{fig5}, upper line). This contrasts with the spectrum of the as-prepared solid (see Figure \ref{fig5}, lower line) and suggests that polymerization has significantly reduced the freedom with which the carbons in C$_{60}$ vibrate. Consistent with $^{13}$C NMR spectroscopic data, the lack of red-shift of the pp-mode to $\sim$1458 cm$^{-1}$ clearly indicates that the polymer does not adopt the [2+2] cycloadduction mode as reported for pure C$_{60}$ thin films \cite{ref17}. Moreover, SEM, HRTEM and SAED studies on the polymerized nanowires (see Figure \ref{fig6}) indicate that the (-C$_{60}$TMB-)$_{n}$ nanopolymers have retained both the crystal morphology and structure of their parent nanowires, \cite{ref29,ref30} reinforcing the view that the polymerization is through a solid-state topochemical event.

\subsection{Favorable Polymerization Pathways}

Detailed investigations of the polymerization reaction mechanism and the nature of the associated bonding mode have been undertaken theoretically. Two possible polymerization scenarios have been considered: the first involves two C$_{60}$ molecules and a TMB molecule and utilizes two methylene bridges; the second involves three C$_{60}$ molecules and a TMB and uses all the three possible linkages. The product predicted from the reaction between two C$_{60}$ molecules and one TMB molecule is a molecular complex, C$_{60}$XC$_{60}$, where X denotes a residue derived from the TMB molecule. There are two possibilities regarding the reaction pathway:

\begin{equation}
{\rm{C}}_{60}  + {\rm{TMB}} + {\rm{C}}_{60}  \to {\rm{C}}_{60} {\rm{ - TMB}}^{{\rm{ - 2H}}} {\rm{ - C}}_{60}  + {\rm{H}}_2  + \Delta E_1
\label{eq15}
\end{equation}

\begin{equation}
{\rm{C}}_{60}  + {\rm{TMB}} + {\rm{C}}_{60}  \to {\rm{HC}}_{60} {\rm{ - TMB}}^{{\rm{ - 2H}}} {\rm{ - C}}_{60} {\rm{H}} + \Delta E_2
\label{eq16}
\end{equation}

\noindent
where TMB$^{\rm{ - 2H}}$ represents the TMB molecule with two hydrogen atoms removed and $\Delta E_1$ and $\Delta E_2$ are the corresponding reaction enthalpies.

\begin{figure}[tbhp]
\includegraphics[width=14cm,clip]{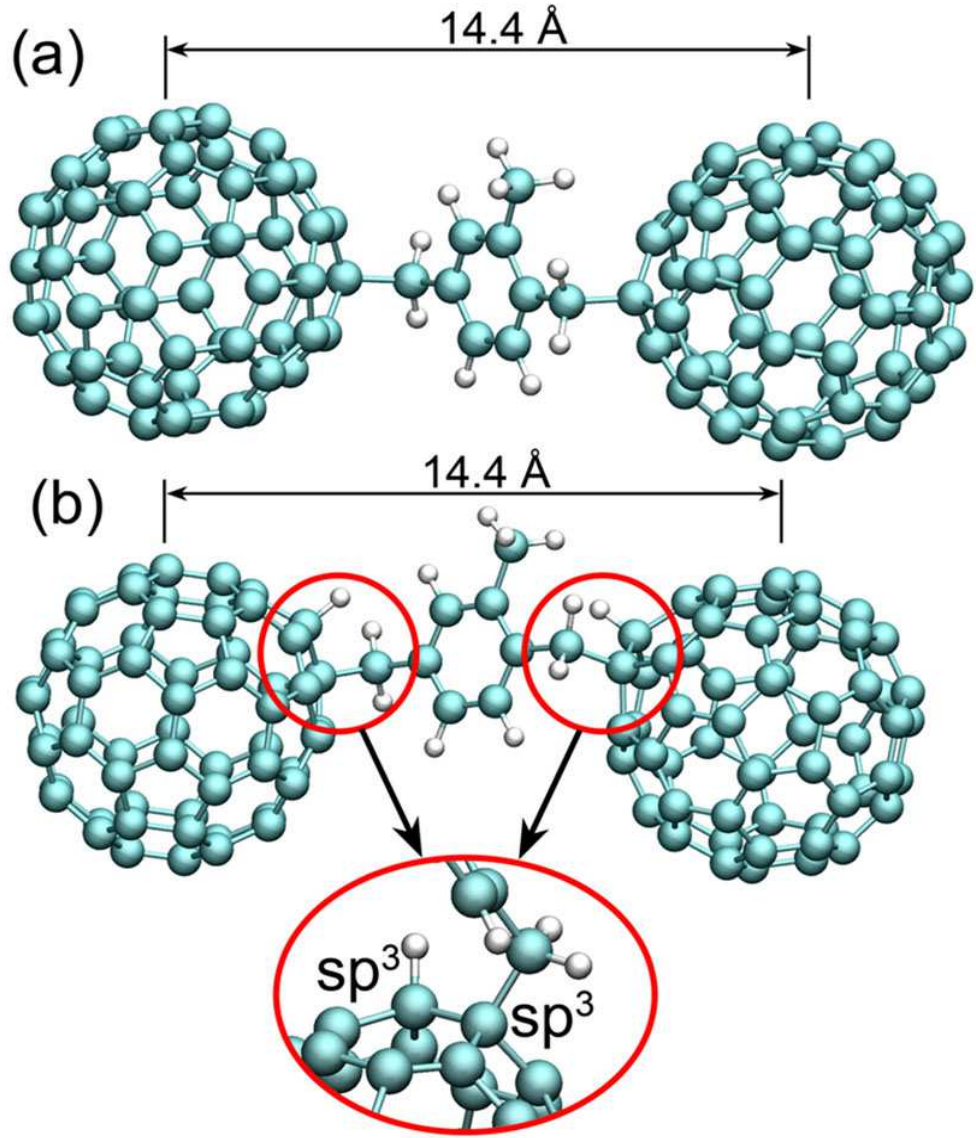}
\caption{Computed structures of the reaction products C$_{60}$-TMB$^{\rm{ - 2H}}$-C$_{60}$ (a) and HC$_{60}$-TMB$^{\rm{ - 2H}}$-C$_{60}$H (b), corresponding to reactions (\ref{eq15}) and (\ref{eq16}) in the text. The distances between the fullerenes are indicated and they are the values calculated using the B3LYP/6-21G method. The inset shows the sp$^{3}$ hybridization of the fullerene carbon atoms resulting from the formation of the covalent bonds with TMB$^{\rm{ - 2H}}$.}
\label{fig7}
\end{figure}

In reaction (\ref{eq15}), two C-H bonds in the TMB molecule (likely at the 1- and 4-methyl positions) are broken and two C-C$_{60}$ bonds are formed. As a consequence, an H$_2$ molecule is released. Reaction (\ref{eq16}) occurs in a similar way, but instead of forming a H$_2$ molecule, the released H-atoms bond to either fullerene. The structures of the computed  (by the {\it ab initio} B3LYP/6-21G method) \cite{ref36} reaction products, C$_{60}$-TMB$^{\rm{ - 2H}}$-C$_{60}$ and HC$_{60}$-TMB$^{\rm{ - 2H}}$-C$_{60}$H, are shown in Figure \ref{fig7}a and \ref{fig7}b, respectively, along with their associated distances between the two fullerenes \cite{ref36}.

To quantify $\Delta E_1$ and $\Delta E_2$, we first optimized the structures of all the reaction partners and calculated their energies using the methods described in section \ref{sec3}. The total energies of the molecules involved in the reactions are summarized in Table \ref{tab2}. The calculated enthalpies are as follows: $\Delta E_1$=59.8 kcal mol$^{-1}$, $\Delta E_2$=11.7 kcal mol$^{-1}$. Note that the B3LYP/STO-3G method gives somewhat different values for the reaction enthalpies. This method is less accurate than the B3LYP/6-21G method, and can, therefore, be only used for a qualitative analysis of the system \cite{ref36}.

\begin{table}[t]
\caption{Total energies of the molecules involved in the analyzed polymerization reactions and the enthalpies ($\Delta E$) of reactions (\ref{eq15}), (\ref{eq16}) and (\ref{eq17}) calculated using {\it ab initio} B3LYP/6-21G and B3LYP/STO-3G methods. The energies are given in atomic units.* The superscripts (a) and (b) indicate the isomers shown in Figure \ref{fig8}. The enthalpies in brackets correspond to the values calculated using the B3LYP/STO-3G method, while the values without brackets were calculated using the B3LYP/6-21G method. \emph{(1 a.u. = 27.2116 eV = 627.499 kcal mol$^{-1}$)}}
\label{tab2}
\begin{center}
\begin{ruledtabular}
\begin{tabular}{|lccc|}
  Molecules & Total energy (a.u.) & Total energy (a.u.)  & $\Delta E$ (kcal mol$^{-1}$) \\
            & B3LYP/6-21G         & B3LYP/STO-3G         &                              \\
  \hline
 1,2,4-TMB                        & -349.8621    &  -345.9655  &  -      \\
 C$_{60}$                         & -2283.9917   &  -2258.2208 &  -      \\
 H$_2$                            & -1.1706      &   -1.1655   &  -      \\
 C$_{60}$-TMB$^{\rm{ - 2H}}$-C$_{60}$    &  -4916.5795  & -4862.4656  & +59.8   \\
                                  &              &             & (+51.0) \\
 HC$_{60}$-TMB$^{\rm{ - 2H}}$-C$_{60}$H  &  -4917.8642  & -4861.1603  & -11.7   \\
                                  &              &             & (-36.71)\\
 3C$_{60}$TMB$^{(a)}$             &  -           & -7120.7144  & -       \\
                                  &              &             & (-53.6) \\
 3C$_{60}$TMB$^{(b)}$             &    -         & -7120.7144  &  -      \\
                                  &              &             & (-55.1) \\
\end{tabular}
\end{ruledtabular}
\end{center}
\end{table}

The fact that reaction (\ref{eq15}) is endothermic can be understood in terms of the changes in electron configuration for carbon atoms in C$_{60}$ following the reaction. Every carbon atom in a fullerene has three covalent bonds with its neighbors, two of which are single, and one of which is double. The carbon atoms in a fullerene therefore exhibit sp$2$ hybridization, but the framework is slightly distorted because of the surface curvature. To attach a TMB molecule to a fullerene, it is necessary to break a double bond in the carbon shell, leading to the formation of two unsaturated carbon atoms. In reaction (\ref{eq15}) the TMB$^{\rm{ - 2H}}$ complex caps one of these carbon atoms in each fullerene, rendering the carbon sp$^{3}$ hybridized but leaving the neighboring carbon unsaturated~--~an energetically unfavorable scenario. In contrast, reaction (\ref{eq16}) is exothermic because the hydrogen atoms released from the TMB molecule cap the neighboring unsaturated carbon atoms in each of the C$_{60}$ fragments. Thus, both carbon atoms that have undergone reaction in either fullerene have become sp$^3$ hybridized (as illustrated by the inset of Figure \ref{fig7}b).

This result is consistent with our elemental analysis for a polymerized sample where parallel analyses yield an average hydrogen content of 1.35 wt\%, consistent with the molar ratio of $\sim$1:1 for 1,2,4-TMB over C$_{60}$ in the solid, as measured previously by thermal gravimetric analysis (TGA) \cite{ref29}. Reaction (\ref{eq16}) has some similarities to the Prato and Bingel reactions widely known in fullerene chemistry \cite{ref46}. In these reactions an organic molecule reacts with a double bond in C$_{60}$ and forms a ring peripheral to the fullerene superstructure. The calculated reaction enthalpy suggests that reaction (\ref{eq16}) is energetically favorable, and therefore is likely to occur.

\begin{figure}[tb]
\includegraphics[width=16cm,clip]{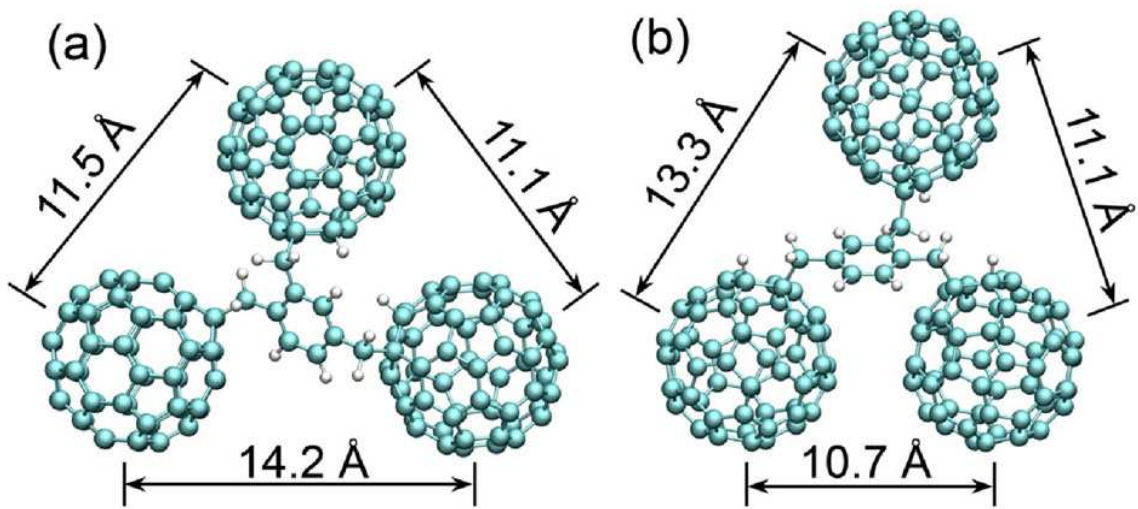}
\caption{Calculated structures of two low energy isomers of the 3C$_{60}$TMB complex as the product of reaction (\ref{eq17}) in the text. The distances between fullerene centers correspond to the geometries optimized at the B3LYP/STO-3G level of calculation.}
\label{fig8}
\end{figure}

A second and more complex polymerization pathway that can be envisaged involves three fullerenes and one TMB. By analogy to equation (\ref{eq16}), the reaction can be written as:

\begin{equation}
{\rm{C}}_{60}  + {\rm{TMB}} + {\rm{C}}_{60}  + {\rm{C}}_{60}  \to 3{\rm{C}}_{60} {\rm{TMB}} + \Delta E_3
\label{eq17}
\end{equation}

\noindent
The possible geometrical arrangements of two low-energy isomers of 3C$_{60}$TMB (B3LYP/STO-3G method) are shown in Figure \ref{fig8}. These structures are constructed in a similar way to the 2C$_{60}$TMB complex, but in this case three C-H bonds have been broken in the TMB unit, and the resulting 3C$_{60}$TMB complex adopts a triangular configuration. The total energies calculated for two optimized isomeric structures of the 3C$_{60}$TMB are shown in Table \ref{tab2}, corresponding to the two complexes shown in Figure \ref{fig8}. The calculated enthalpies are $\Delta E_3$=-53.64 kcal mol$^{-1}$ and $\Delta E_3$=-55.11 kcal mol$^{-1}$, for the product isomers (a) and (b), respectively. In either case, it is clear that reaction (\ref{eq17}) is exothermic, and thus energetically favorable.

\subsection{Structural Calculations and Analysis}

\begin{figure}[htbp]
\includegraphics[width=16cm,clip]{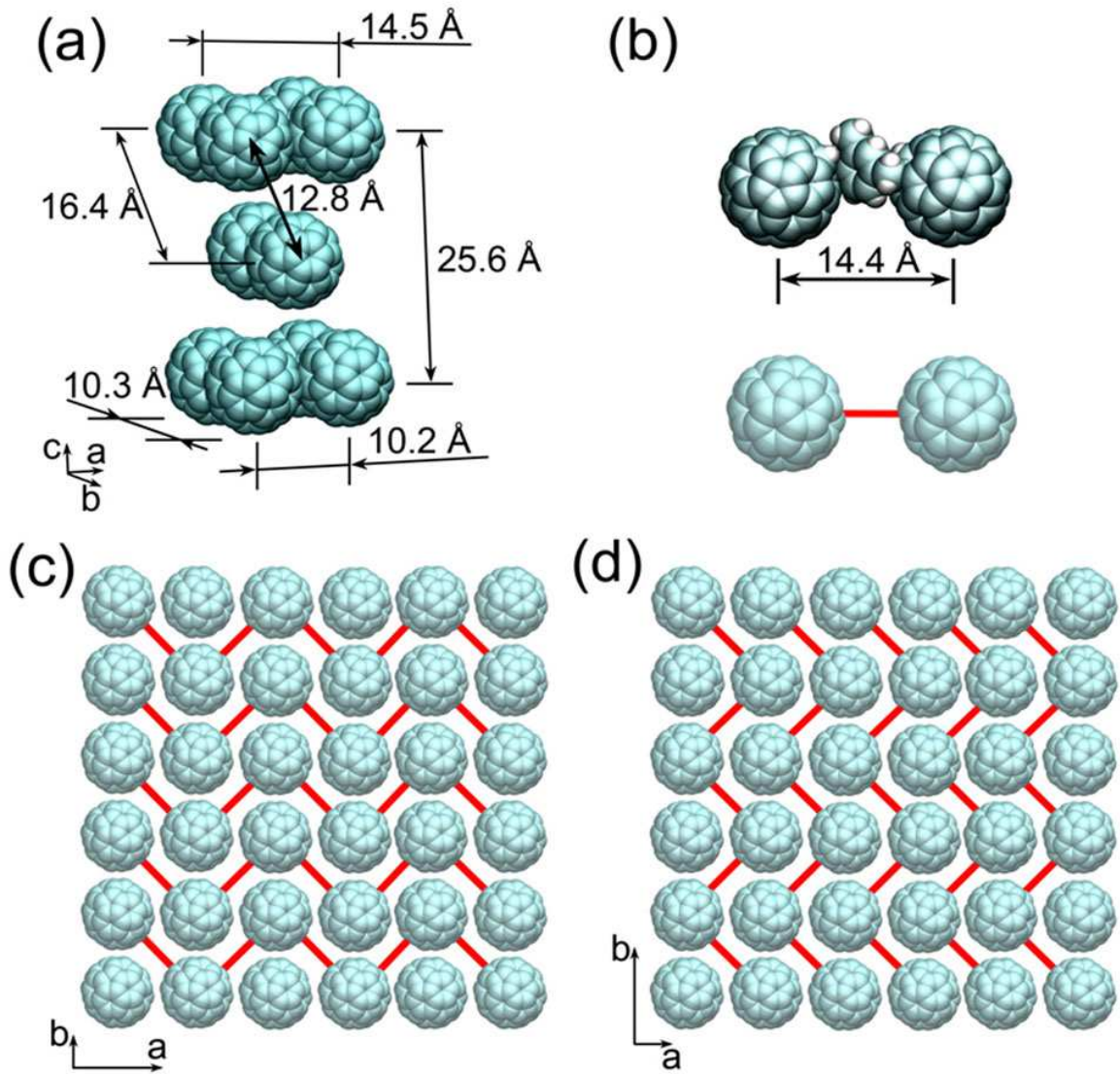}
\caption{Possible polymerization scenarios of the C$_{60}$TMB 1D nanocrystals. In (a) we show the geometry of the unit cell (the TMB molecules embedded in the unit cell are not shown). The distances between the centers of the fullerenes are indicated. The geometry of a C$_{60}$-TMB-C$_{60}$ molecule is shown in (b). The distance between the two fullerenes in this case is 14.4 \AA, very close to the dimensions of the unit cell. Plots (c) and (d) demonstrate the cross section of a nanowire, viewed along the c-axis, and illustrate two possible linking schemes between the fullerenes. The bars between the fullerenes indicate the TMB molecules. The polymerization can occur along either the a- (plot c) or the b-axis (plot d) of a nanowire \cite{ref29,ref30}.}
\label{fig9}
\end{figure}

In order to understand how the 1,2,4-TMB molecules bond to fullerenes in the nanowires, we have studied geometrical constraints of the unit cell of the crystal lattice. Figure \ref{fig9}a shows the geometry of such a unit cell and its dimensions, based on the experimental data \cite{ref29,ref30,ref47}. For simplicity, here we do not show the embedded TMB molecules. As discussed before, a TMB molecule may link two fullerenes together, and if this mode is repeated periodically, the crystalline nanowire would turn into a nanopolymer. However, this is only possible if a TMB fits into the space between two fullerenes without significantly disturbing the unit cell structure.

In Figure \ref{fig9}b we show a C$_{60}$-TMB-C$_{60}$ complex, where two fullerenes are covalently linked by a 1,2,4-TMB molecule. In this case the distance between the two fullerenes is 14.4 \AA, very close to one of the inter-fullerene spacings measured for the unit cell (14.5 \AA, Figure \ref{fig9}a). The linkage is thus highly likely to occur in the unit cell and the C$_{60}$-TMB-C$_{60}$ complex is likely to be a possible building block for the polymerization. Figures \ref{fig9}c and \ref{fig9}d illustrate two possible linking schemes between the fullerenes, where the bars between fullerenes represent TMB molecules. Based on the preferential crystal growth directions studied previously \cite{ref29,ref30}, this polymerization may occur along either the a- (see Figure \ref{fig9}c) or the b-axis (see Figure \ref{fig9}d), with the linkages corresponding to the two schemes as shown in the figures.

\begin{figure}[t]
\includegraphics[width=16cm,clip]{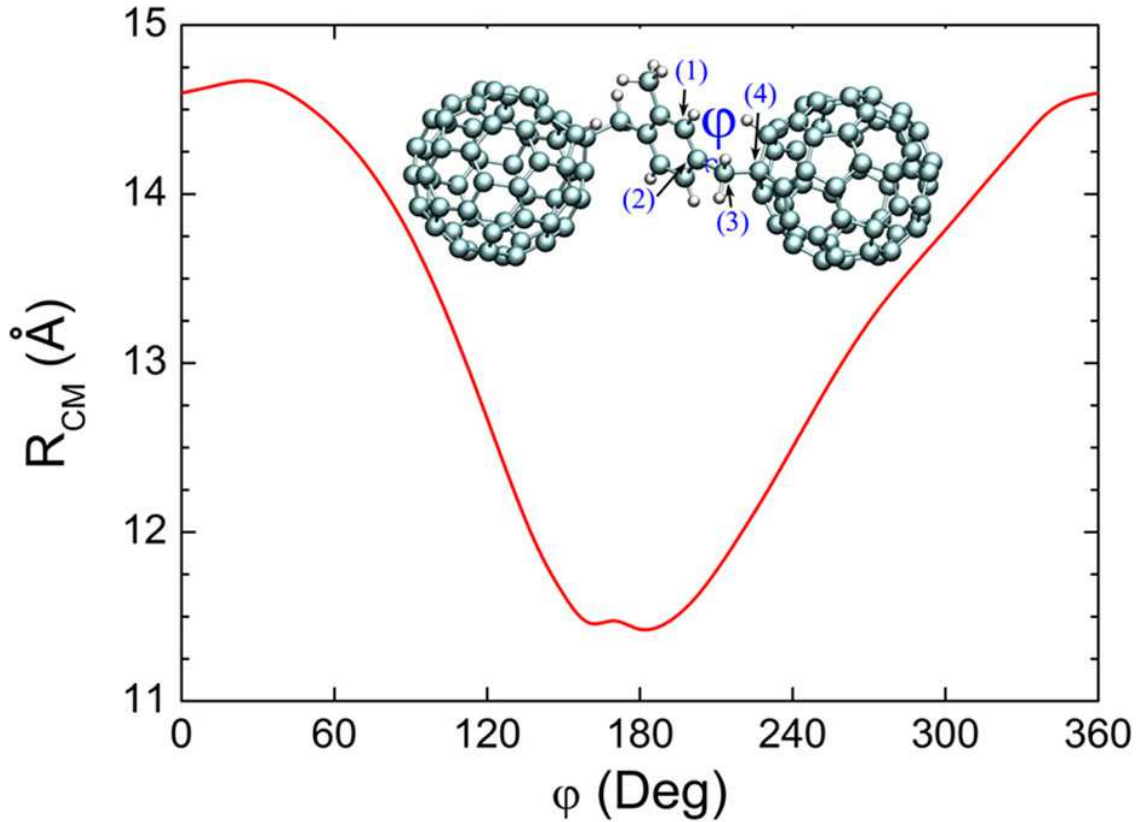}
\caption{Distance between the centres of two fullerenes in a C$_{60}$-TMB-C$_{60}$ complex as a function of angle $\varphi$, which is defined as the angle between the planes formed by the atoms (1)-(2)-(3) and (2)-(3)-(4), as shown in the inset.}
\label{fig10}
\end{figure}

It is important to stress that a C$_{60}$-TMB-C$_{60}$ complex has several stable isomeric configurations with different C$_{60}$-C$_{60}$ distances. To illustrate this fact, in Figure \ref{fig10} we show how the the distance between two fullerenes in a C$_{60}$-TMB-C$_{60}$ complex depends on an angle $\varphi$, which is defined as the separation degree between the planes formed by atoms (1)-(2)-(3) and atoms (2)-(3)-(4), as shown in the Figure. As the bonds connecting two C$_{60}$ units and a TMB$^{\rm{ - 2H}}$ are not collinear, the center-to-center distance changes upon rotation of the fullerenes. As can be seen from Figure \ref{fig10}, such a distance varies from 11.5 \AA\ to 14.5 \AA, and match the distances between fullerenes in the unit cell (as indicated in Figure \ref{fig9}a).

\begin{figure}[t]
\includegraphics[width=16cm,clip]{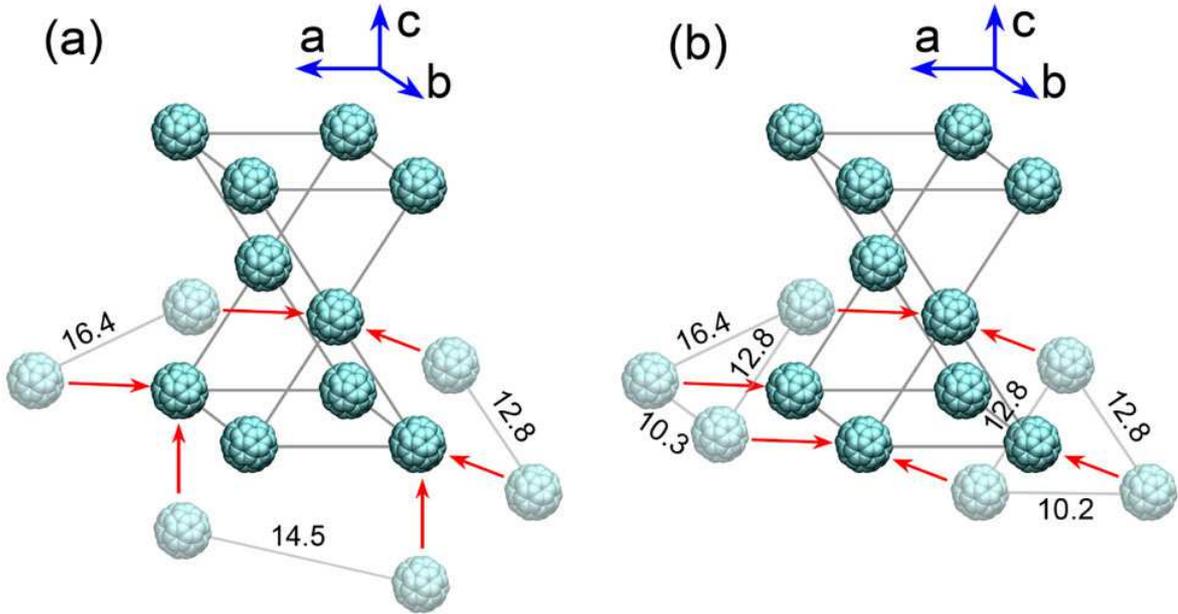}
\caption{Possible fitting sites of the 2C$_{60}$TMB complex (a), and the 3C$_{60}$TMB complex (b), in the unit cell of the parent nanowire. Distances are given in {\AA}ngstroms and correspond to experimental data \cite{ref29}.}
\label{fig11}
\end{figure}

Based on this geometrical analysis, we have noted that there appear to be three possibilities for fusing a TMB$^{\rm{ - 2H}}$ molecule in between two fullerenes in the unit cell, as illustrated by the three sets of C$_{60}$ dimers marked in Figure \ref{fig11}a. Except these possibilities, the distances between any other two fullerenes are either too small or too large and are thus incompatible with the requirement for a topochemical polymerization to occur. Note that although the distance 16.4 \AA in the unit cell is somewhat larger than the distances shown in Figure \ref{fig10}, the related fusion mode is still possible because the distance between fullerenes in a C$_{60}$-TMB-C$_{60}$ complex may further change at certain conditions. Some deviations are expected because the C$_{60}$-TMB-C$_{60}$ complex was optimized in vacuo without accounting for other related constituents in the polymer. Also, the complex has many isomers because in the crystal unit cell a fullerene has many different atomic sites to which a TMB$^{\rm{ - 2H}}$ molecule may attach.

Similarly, Figure \ref{fig11}b shows two possible fitting sites for a 3C$_{60}$TMB complex in the unit cell. This structural analysis clearly demonstrates that a 1,2,4-TMB molecule can easily stitch two fullerenes together within a nanowire without disturbing the lattice structure. This conclusion is in excellent agreement with the experimental studies with SEM, HRTEM and SAED, as described earlier.

\subsection{Theoretical Analysis of the $^{13}$C-NMR Spectra}

Our experimentally measured chemical shifts are supported by the calculated values as shown by Figure \ref{fig3}e. To do the calculation, we have applied the {\it ab initio} B3LYP/6-31G(d) method to pristine C$_{60}$ (dotted line, red fill), pure 1,2,4-TMB (dashed line, green fill) and the C$_{60}$-TMB-C$_{60}$ complex (solid line, blue fill). Since many of the calculated chemical shifts are degenerate, Lorentzian broadening for each line is assumed:

\begin{equation}
F(\delta ) = \frac{1}{\pi }\sum\limits_{i = 1}^N {\frac{{{\raise0.7ex\hbox{$1$} \!\mathord{\left/
 {\vphantom {1 2}}\right.\kern-\nulldelimiterspace}
\!\lower0.7ex\hbox{$2$}}\sigma }}{{\left( {\delta  - \delta _i } \right)^2  + \left( {{\raise0.7ex\hbox{$1$} \!\mathord{\left/
 {\vphantom {1 2}}\right.\kern-\nulldelimiterspace}
\!\lower0.7ex\hbox{$2$}}\sigma } \right)^2 }}}
\label{eq18}
\end{equation}

\noindent
Here $\delta _i$  is the calculated chemical shift, and $\sigma  = {\raise0.7ex\hbox{$2$} \!\mathord{\left/
 {\vphantom {2 \pi }}\right.\kern-\nulldelimiterspace}
\!\lower0.7ex\hbox{$\pi $}}$  is the width of peaks at half maximum intensity. The calculated $^{13}$C-NMR spectra are plotted on a logarithmic scale for better visualization.

The calculated chemical shifts for pure C$_{60}$ and 1,2,4-TMB are in good agreement with experiment. The observed signal for C$_{60}$ at 143.3 ppm is only 4 ppm higher than the calculated, and the calculated shift for the ring carbons of TMB is only 7-8 ppm lower than the observed. Also, the calculated values for the methyl groups perfectly reproduce the measured data (within an error of less than 1.0 ppm). These comparisons enable us to conclude that the accuracy of the calculation is on the order of $\sim$5-8 ppm, which is sufficient for an unambiguous characterization of the calculated NMR peaks in this work.

In the previous section we suggested that the C$_{60}$-TMB-C$_{60}$ complex was a building block in the polymerization. The calculated $^{13}$C-NMR spectrum for the complex strongly supports this view by virtue of its excellent agreement with the experimental result. Crucially, the emergence of four peaks in the range of 45-80 ppm, consistent with the observed set of broad peaks at 55-85 ppm, corresponds to the chemical shifts of the carbon atoms (2), (3) and (4), respectively, as marked in Figure \ref{fig3}f. These three atoms participate in the bonds formed between C$_{60}$ and 1,2,4-TMB molecules in the polymerized network.

\begin{figure}[htb]
\includegraphics[width=16cm,clip]{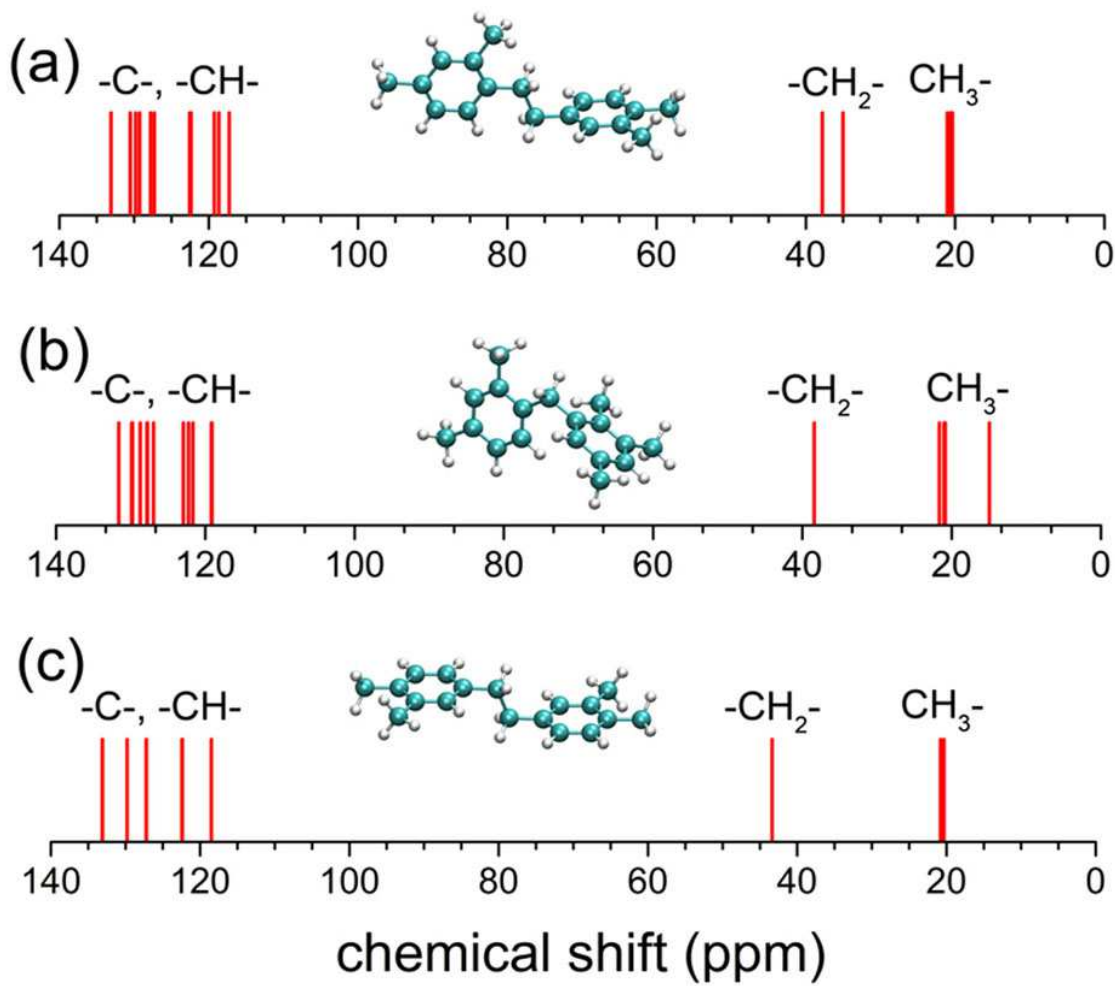}
\caption{$^{13}$C-NMR spectra calculated for three isomeric states of the TMB-TMB dimer using the {\it ab initio} B3LYP/6-31G(d) method. The geometries of the molecules used in the calculation are shown in the inset. Labels indicate the chemical shifts of different types of carbon atoms in the system. Note that the chemical shifts in (c) are much more degenerate than (a) and (b) due to the enhanced symmetry of the studied isomer.}
\label{fig12}
\end{figure}

The observed peak at 30.4 ppm can be interpreted as the response from methylene carbons cross-linking adjacent 1,2,4-TMB molecules due to polymerization. This view is strongly supported by literature report of the structurally closely-related framework of multi-substituted diphenylmethane (see Table \ref{tab1}) \cite{ref43,ref44,ref45}. Within the accuracy of the calculation, it is also consistent with our calculations for TMB-TMB dimers, which predict the corresponding chemical shift to be in the range of 28-43 ppm, depending on the bonding mode adopted by the dimers and the relative orientations of the adjacent TMB molecules involved (see Figure \ref{fig12}). Figure \ref{fig12}a shows an example of a dimer in which the two benzene rings are perpendicular to each other, and the two TMB molecules are bonded together through two methylene carbons corresponding to the ring positions of 1- and 4-methyl, respectively. In this case, the two methylenes yield two closely placed peaks at 35.7$\pm$7 ppm and 37.8$\pm$7 ppm. Figure \ref{fig12}b shows another example of a dimer in which the two TMB molecules are linked by only one methylene carbon which exhibits a chemical shift of 38.4 ppm. Figure \ref{fig12}c shows the third example of such a dimer isomer, where the two TMB molecules are linked by two identical methylene carbons but the two benzene rings are parallel to each other. In this case the chemical shift calculated is 43.4 ppm. These analyses suggest that the chemical shift of the methylene carbons is sensitive to the surrounding atoms, and that the peak observed at 30.4 ppm hint at some specific bonding mode and/or relative orientations between TMB molecules in the polymeric chains. Moreover, the enthalpy calculated for such a dimerization reaction using the B3LYP/6-21G method is -19.2 kcal/mol, indicating that the dimerization reaction for polymerization is energetically favorable.

\section{Conclusion and perspective}

We have demonstrated for the first time an approach to the synthesis of a C$_{60}$-based nanowire polymer and established the chemical bonding mode involved in the polymeric chains based on both experimental measurements and theoretical calculations. Importantly, the material adopts a crystalline 1D nanostructure which resembles carbon nanotubes in shape and other important conjugated polymers in structure. Since the material does not contain any metal but is simply composed almost entirely of carbon (while it contains hydrogen, the content is only 1.4 wt \%), it suggests biological compatibility and it is, perhaps, even more attractive than carbon nanotubes for bio-applications. In addition, the material has further important potential for applications in photo-electrical devices because of the intrinsically large magnitude of the nonlinear optical response of C$_{60}$ and the excellence of its photoinduced charge transfer properties. Considering all these, we believe that this work represents a step toward true applications of C$_{60}$ in nanotechnology by the ability of processing commercially available raw C$_{60}$ powder into a one-dimensional, crystalline, and covalently-bonded fullerene nanopolymer.

We consider that applications of the reported nanopolymer may be facilitated by a wet-chemical approach through surface modification of the material using the rich chemistry of fullerene developed over the last 20 years. Since the nanopolymer is insoluble in common solvents, such surface modification or functionalization should be possible to achieve in either an aqueous or an organic solution without destructing its solid-state structure. Such a wet approach would benefit from low-cost processing, the need for only simple apparatus and the possibility of scaling-up to the industrial level. Moreover, the nanopolymer itself not only provides an example of phase transition of the parent nanowire driven by forming and breaking covalent bonds, but also illustrates the enduring significance of the original fullerene concept and its versatility as applied to new fullerene-related nanostructures. Finally, the host (C$_{60}$) and guest (1,2,4-TMB) nature of the polymerization suggests a general host-guest route to the synthesis of new types of fullerene-based nanopolymers constructed by different organic monomers and fullerenes.

\section{Acknowledgement}
We thank support from the European Commission project (NoE EXCELL NMP3-CT-2005-515703). The assistance of Dr. Wuzong Zhou, School of Chemistry, St. Andrews University, for carrying out TEM analysis is acknowledged. Computer simulations were carried out at the Frankfurt Center for Scientific Computing.

\end{document}